\documentclass[aps,prd,twocolumn,superscriptaddress,nofootinbib,10pt]{revtex4-1}
\usepackage{amsmath,amssymb,bm,latexsym,acronym,float,soul,nicefrac,verbatim,cancel}
\usepackage{multirow,dcolumn,longtable,footnote,appendix,graphicx,color,subfigure}
\usepackage[colorlinks,linkcolor=blue,citecolor=blue,urlcolor=blue ]{hyperref}
\usepackage[normalem]{ulem}
\newcommand{\TRC}{MOE Key Laboratory of TianQin Mission, TianQin Research Center for Gravitational Physics $\&$ School of Physics and Astronomy, Frontiers Science Center for TianQin, Gravitational Wave Research Center of CNSA, Sun Yat-sen University (Zhuhai Campus), Zhuhai 519082, China}

\allowdisplaybreaks
\newcommand{\be}{\begin{equation}}
\newcommand{\ee}{\end{equation}}
\newcommand{\bea}{\begin{eqnarray}}
\newcommand{\eea}{\end{eqnarray}}
\newcommand{\nn}{\nonumber}
\newcommand{\cK}{{\cal K}}
\newcommand{\cO}{{\cal O}}
\newcommand{\cL}{{\cal L}}

\newcommand{\cQ}{{\cal Q}}

\newcommand{\pd}{\partial}
\newcommand{\td}{\tilde}
\newcommand{\wtd}{\widetilde}

\newcommand{\tdh}{{\td{h}}}

\begin{document}

\title{\bf Separating the linearized Einstein equations in the background of a Kerr black hole}

\author{Jianwei Mei}
\email{Email: meijw@sysu.edu.cn}
\affiliation{\TRC}

\begin{abstract}
The advancement in gravitational wave detection has made it possible to study the nonlinear effects in black hole perturbations, the modeling of which requires the full knowledge of the linear order perturbation of the metric.
For the most interesting Kerr background, the conventional practice is to solve the Teukolsky equations for a Weyl scalar, and then the perturbation of the metric components is obtained using an elaborated metric reconstruction scheme, which, however, comes with unwanted restrictions.
So it is desirable to separate the linearized Einstein equations in Kerr directly, i.e., without using metric reconstruction.
In this work, I present a method that aims to help achieve this goal.\\ \\
{\bf Keywords:} black hole perturbation, linearized Einstein equations, separation of variables, Killing-Yano tensor, symmetry
\end{abstract}

\maketitle

%\keywords{}

%\pacs{04.20.Cv,04.50.Kd,04.80.Cc,04.80.Nn}

%%%%%%%%%%%%%%%%%%%%%%%%%%%%%%%%%%%%%%%%%%%%%%%%%%%%%%%%%%%%%%%%
\acrodef{GR}{general relativity}
\acrodef{GW}{gravitational wave}
\acrodef{EMRI}{Extreme Mass Ratio Inspiral}
\acrodef{MBHB}{massive black hole binary}
\acrodef{BHPT}{Black hole perturbation theory}
%%%%%%%%%%%%%%%%%%%%%%%%%%%%%%%%%%%%%%%%%%%%%%%%%%%%%%%%%%%%%%%%

\section{Introduction}

Black hole is one of the most intriguing predictions of \ac{GR}.
Observations indicate that almost every galaxy has a massive black hole at its center \cite{Kormendy1995} and galaxies like the Milky Way may contain millions of stellar mass black holes \cite{Elbert:2017sbr}.
Given the difficulty for astrophysical black holes to maintain significant amount of electric charges \cite{Gibbons:1975kk,Ghosh:2022kit}, the Kerr hypothesis \cite{Carter:1971zc,Bekenstein:1996pn,Chrusciel:2012jk,Cardoso:2016ryw} predicts that all stationary black holes in the Universe are described by the Kerr metric \cite{Kerr:1963ud},
%%%
\bea ds^2&=&-dt^2+H\Big(\frac{dr^2}{X}+\frac{dx^2}Y\Big)+(r^2+a^2)Yd\phi^2\nn\\
&&+\frac{2Mr(dt-aYd\phi)^2}{H}\,, \label{metric.kerr}\eea
%%%
where $X=r^2+a^2-2Mr\,$, $Y=1-x^2$, $H=r^2+a^2x^2\,$, and the coordinates are Boyer-Lindquist.

The breakthrough in \ac{GW} detection has opened the door to testing the nature of black holes with unparalleled precisions \cite{Barack:2018yly,Luo:2025ewp,Berti:2025hly}.
In fact, the expected precisions are so high that it becomes possible to probe the nonlinear effects in the signals, such as the second order self-force effect in \ac{EMRI} signals \cite{Poisson:2011nh,Barack:2018yvs,Pound:2021qin} and the quadratic modes in black hole ringdowns \cite{Cheung:2022rbm,Mitman:2022qdl,Qiu:2023lwo,Shi:2024ttu,Yi:2024elj,Khera:2024bjs,Lagos:2024ekd,Berti:2025hly}.
The detection of such nonlinearity has become an important scientific objective for both the space-based \cite{Luo:2025ewp,LISA:2022kgy} and the next-generation ground-based \ac{GW} detectors \cite{Yi:2024elj,Khera:2024bjs,Lagos:2024ekd,Kalogera:2021bya,Reitze:2019iox,Punturo:2010zz}.

To model the nonlinear effects in the \ac{GW} signals, it is necessary to have full knowledge of the linear order perturbation of the metric.
This is usually done by starting with Teukolsky's master equation, which is a second order partial differential equation for a function $\Psi(t,r,x,\phi)$ that encodes the perturbation of the Kerr metric by fields with various spins, $|s|=0,\frac12,1,2$ \cite{Teukolsky:1972my,Teukolsky:1973ha,Teukolsky:2014vca}.
Using the ansatz $\Psi(t,r,x,\phi)=e^{i(wt-m\phi)}R(r)P(x)\,$, Teukolsky's master equation can be reduced to two ordinary differential equations,
%%%
\bea R''&=&\frac1X\Big\{\td\lambda+4iswr-\frac1X[am-(r^2+a^2)w]\nn\\
&&\qquad\times[am-(r^2+a^2)w-2is(r-M)]\Big\}R\nn\\
&&-\frac{2(1+s)(r-M)}XR'\,,\nn\\
P''&=&-\frac1Y\Big[\td\lambda+s(1+s)+\frac{2sx(m+awY)-s^2}Y\nn\\
&&\qquad-\frac{(m-awY)^2}Y\Big]P+\frac{2x}YP'\,,\label{eom.master}\eea
%%%
where $\td\lambda$ is a separation constant.

For the metric perturbations, one may take $\Psi=\psi_0$ with $s=2$ and $\Psi=(r-iax)^4\,\psi_4\,$ with $s=-2$, where $\psi_0$ and $\psi_4$ are the extreme Weyl scalars.
One can solve for $\psi_0$ or $\psi_4$ from the ansatz of $\Psi$ and \eqref{eom.master}, and then reconstruct the metric perturbations through the procedure initially developed in \cite{Chrzanowski:1975wv,Kegeles:1979an,Wald:1978vm,Stewart:1978tm}.
However, the metric reconstruction procedure is computationally tedious and has obvious limitations, such as the necessity to use the radiation gauge \cite{Dolan:2021ijg} and to solve an inversion problem involving fourth order differential equations in the intermediate steps \cite{Wald:1978vm,Lousto:2002em,Berens:2024czo}.
The need for metric reconstruction has added to the difficulty of formulating non-linear perturbations of the Kerr black hole, see, e.g.,  \cite{Merlin:2016boc,Green:2019nam,Loutrel:2020wbw,Toomani:2021jlo,Aly:2023qzk,Wardell:2021fyy,Ripley:2020xby,Dolan:2023enf,Sberna:2021eui,Yang:2014tla,Green:2022htq,Spiers:2024src,Chung:2023wkd,Hollands:2024iqp}.
So, it will be of great importance to find a generic metric scheme that can be used to separate the linearized Einstein equations in Kerr directly, i.e., without using metric reconstruction.
There have been some attempts in this direction in the past, but they all come with significant restrictions, such as by focusing on the near-horizon extremal Kerr background \cite{Chen:2017ofv} or taking the small rotation limit \cite{Franchini:2023xhd}.

In this paper, I try to approach the problem by following the logically more straightforward route, i.e., by using the symmetries of the linearized Einstein equations.
In general, one expects that the linear order perturbation of the metric can be decomposed as
%%%
\bea&&h_{\mu\nu}(t,r,x,\phi)=\sum_{\lambda wm}A_{\lambda wm} h^{(\lambda wm)}_{\mu\nu}(t,r,x,\phi)\,,\nn\\
&&h^{(\lambda wm)}_{\mu\nu}(t,r,x,\phi) = f^{(\lambda wm)}_{\mu\nu}(r,x)e^{i(wt-m\phi)}\,,\label{decomp}\eea
%%%
where $h^{(\lambda wm)}_{\mu\nu}$ is the function basis used for the decomposition (each is called a mode), and $A_{\lambda wm}$ is the decomposition constant for each mode.
The factor $e^{i(wt-m\phi)}$ in $h^{(\lambda wm)}_{\mu\nu}$ is due to the time translation and rotation symmetry of the equations, and $\lambda$ stands for a new mode index that is expected to be the eigen value of some conserved operator that is yet to be found.
So the questions are:
\begin{itemize}
\item Does such an operator exist?
\item Can it be used to help separate the linearized Einstein equations in Kerr?
\end{itemize}

In this paper, I explicitly construct such an operator and show with concrete examples that it can be used to help separate the linearized Einstein equations in Kerr.
Two explicit solutions of $f^{(\lambda wm)}_{\mu\nu}$ have been found showing how each metric component depends on the functions $R(r)$ and $P(x)$, which is the key to reduce the linearized Einstein equations in Kerr to the two ordinary differential equations in \eqref{eom.master}.

In the following, all covariant derivatives are defined with the background metric (\ref{metric.kerr}),
a prime on any function always means a derivative with respect to the corresponding argument of the function,
and the coordinates are numbered according to the sequence $\{t,r,x,\phi\}\,$.

\section{Separating variables with a symmetry operator}

Consider the general problem of separating the variables for the wave equation,
%%%
\be \wtd\Box \Psi=0\,,\label{eom.Psi}\ee
%%%
where the details of the wave operator $\wtd\Box$ will depend on the nature of the field $\Psi\,$ that it acts on.

The separability of (\ref{eom.Psi}) over the Kerr background is known to be related to the existence of a Killing-Yano tensor \cite{Walker:1970un,Frolov:2017kze},
%%%
\bea k_{\mu\nu}&=&\left(\begin{matrix}0&ax&ar&0\cr -ax&0&0&a^2xY\cr -ar&0&0&r(r^2+a^2)\cr
0&-a^2xY&-r(r^2+a^2)&0\end{matrix}\right)\,,\eea
%%%
which satisfies $\nabla_\mu k_{\nu\rho}+\nabla_\nu k_{\mu\rho}=0\,$. The ``square" of the Killing-Yano tensor, $K_{\mu\nu}=k_{\mu\rho}k_\nu^{~\rho}\,$, satisfies $\nabla_\mu K_{\nu\rho}+\nabla_\nu K_{\rho\mu}+\nabla_\rho K_{\mu\nu}=0\,$. Note $K=K_\alpha^{~\alpha}=2H\,$.

To help separate the variables in (\ref{eom.Psi}), one can construct a symmetry operator, $\cK$, that commutes with $\wtd\Box$,
%%%
\be[\wtd\Box\,,\,\cK]=0\,,\label{eom.cK.com}\ee
%%%
and then the eigen-equation
%%%
\be \cK\Psi=\lambda\Psi\,,\label{eom.eigen}\ee
%%%
will supply the needed separation constant, here denoted by $\lambda$.
This work will focus on $\cK$ being a double-quadratic operator, i.e., one that is quadratic both in the covariant derivative and in the Killing-Yano tensor. Similar to $\wtd\Box\,$, the detail of $\cK$ will also depend on the nature of the field that it acts on.

In the following, I use the scalar and vector wave equations to illustrate how the method can be applied.

\subsection{Scalar}

Separated scalar wave equations over the Kerr background have been explicitly known long ago \cite{Brill:1972xj}.
Here it is used as the simplest example to illustrate the method.
For the purpose of this work, it is enough to consider the massless and source-free scalar field.
The field equation is given by
%%%
\bea\wtd\Box\Phi\equiv\nabla_\mu\nabla^\mu\Phi=0\,.\label{eom.scalar}\eea
%%%
The symmetry operator can be constructed as
%%%
\bea\cK\Phi=-\nabla_\mu(K^{\mu\nu}\nabla_\nu\Phi)\,.\eea
%%%
One can then solve (\ref{eom.scalar}) in combination with the following eigen equation,
%%%
\bea\cK\Phi=\lambda\Phi\,,\label{eom.scalar.eigen}\eea
%%%
which brings an independent separation constant, $\lambda\,$.

Since the Kerr background is stationary and axisymmetric, one can use the ansatz
%%%
\be \Phi(t,r,x,\phi)=\Phi(r,x)e^{i(wt-m\phi)}\,.\label{ansatz0.scalar}\ee
%%%
In this way, the dependence on the variables $t$ and $\phi$ is easily separated, and the problem is reduced to separating the remaining variables $r$ and $x\,$.
The final step is achieved with $\Phi(r,x)=R(r)P(x)\,$, where $R(r)$ and $P(x)$ are governed by (\ref{eom.master}) with $s=0$ and $\td\lambda=\lambda\,$.
By letting $\lambda=Q+(m-wa)^2\,$, one can recover the result of \cite{Brill:1972xj} in the massless and source-free case.

\subsection{Vector}

Separated vector wave equations over the Kerr background have also been known: based on \cite{Teukolsky:1972my}, Chandrasekhar has solved for the vector field in terms of the separated Teukolsky functions \cite{Chandrasekhar:1976mx,Chandrasekhar:1985kt}, and this work has been generalized to type-D backgrounds by Torres Del Castillo \cite{TorresDelCastillo:1988td}.
Lunin has re-derived the separated equations with symmetry considerations and by using a special ansatz \cite{Lunin:2017drx}.
Here the separated equations are re-derived again with the construction of several symmetry operators that commute with the wave operator.
See \cite{Dolan:2019hcw} for more related results.

For the purpose of this work, it is enough to consider the source-free Maxwell equations,
%%%
\bea \wtd\Box A^\nu\equiv\nabla_\mu F^{\mu\nu} =\nabla_\mu\nabla^\mu A^\nu -\nabla_\mu \nabla^\nu A^\mu=0\,. \label{eom.Au}\eea
%%%
Limited to the expressions that are both quadratic in the derivatives and in the Killing-Yano tensor, four combinations that commute with (\ref{eom.Au}) off-shell can be found,
%%%
\bea\cK_1A^\mu&=&A_\alpha k^{\gamma\mu}\nabla_\beta\nabla_\gamma k^{\alpha\beta}
+A_\alpha k_{\beta\gamma}\nabla^\alpha\nabla^\mu k^{\beta\gamma}\nn\\
&&+\frac16A^\mu\nabla_\gamma k_{\alpha\beta}\nabla^\gamma k^{\alpha\beta}\,,\nn\\
\cK_2A^\mu&=&\frac12k_{\beta\gamma}k^{\beta\gamma}\nabla_\alpha\nabla^\mu A^\alpha
+k_{\beta\gamma}\nabla_\alpha A^\alpha\nabla^\mu k^{\beta\gamma}\,,\nn\\
\cK_3A^\mu&=&k_{\beta\gamma}k_\alpha^{~\gamma}\nabla^\alpha\nabla^\beta A^\mu
-k_{\beta\gamma}\nabla^\gamma k^{\alpha\beta}\nabla_\alpha A^\mu\nn\\
&&-A_\alpha k^{\gamma\mu}\nabla_\beta\nabla_\gamma k^{\alpha\beta}
-A_\alpha\nabla^\gamma k^{\alpha\beta}\nabla^\mu k_{\beta\gamma}\nn\\
&&-2\Big(k_{\alpha\gamma}\nabla^\mu k_\beta^{~\gamma}
-2k_{\beta\gamma}\nabla^\mu k_\alpha^{~\gamma}\nn\\
&&\qquad+k^{\gamma\mu}\nabla_\gamma k_{\alpha\beta}\Big)\nabla^\beta A^\alpha\,,\nn\\
\cK_4A^\mu&=&k_{\alpha\gamma}k_\beta^{~\mu}\nabla^\beta\nabla^\gamma A^\alpha
+\Big(k_{\alpha\gamma}\nabla^\mu k_\beta^{~\gamma}\nn\\
&&-k_{\beta\gamma}\nabla^\mu k_\alpha^{~\gamma}
+2k^{\gamma\mu}\nabla_\gamma k_{\alpha\beta} \Big)\nabla^\beta A^\alpha\,.\eea
%%%
It turns out that not all the operators are on equal footing: $\cK_1A\equiv A$ and so is trivial, $\cK_2A$ vanishes in the Lorenz gauge, $\nabla_\mu A^\mu=0\,$, while $\cK_4A$ is the only one that not only is gauge invariant but also commutes with all the other operators:
%%%
\be [\cK_4\,,\,\cQ]=0\,,\quad \cQ\in\{\wtd\Box\,,\,\cK_i\,,\,i=1,2,3\}\,.\ee
%%%
Because of these properties, $\cK_4$ has been chosen to do the separation, i.e., (\ref{eom.Au}) is solved in combination with
%%%
\be \cK_4A^\mu=\lambda A^\mu\,,\label{eom.eigen.Au}\ee
%%%
where $\lambda$ is the separation constant for the vector field. I have not studied the cases when other symmetry operators are used instead.

With the ansatz
%%%
\be A^\mu(t,r,x,\phi)=A^\mu(r,x)e^{i(wt-m\phi)}\,,\label{ansatz1.vector}\ee
%%%
(\ref{eom.Au}) and (\ref{eom.eigen.Au}) can be solved using
%%%
\bea A^t(r,x)&=&-\frac{i\lambda^2ar(r^2+a^2)(m-wa)Y(Z_1+Z_2)}{HX_2Y_2}\,,\nn\\
A^r(r,x)&=&\frac{\lambda a^2(m-wa)^2XY}{HX_2Y_2}\td{R}\td{P}\,,\nn\\
A^x(r,x)&=&\frac{RP}{H}\,,\nn\\
A^\phi(r,x)&=&-\frac{i\lambda^2a^2r(m-wa)Y(Z_1-Z_2)}{HX_2Y_2}\,,\label{sol.Au}\eea
%%%
where
%%%
\bea Z_1&=&\Big[\td{R}-\frac{2aX_4(r^2+a^2)Y-H_2X(m-waY)}{2\lambda a^2r(r^2+a^2)(m-wa)XY} X_2R\Big]\td{P}\nn\\
&&-\frac{xH_2X_2Y_2RP}{2\lambda a^2r(r^2+a^2)(m-wa)^2Y^2}\,,\nn\\
Z_2&=&-\frac{HX_2(m-waY)R}{2\lambda a^2r(r^2+a^2)(m-wa)Y}\nn\\
&&\times \Big[\td{P} -\frac{xY_2P}{(m-wa)(m-waY)Y}\Big]\,,\eea
%%%
and
%%%
\bea\td{R}&=&R'+\frac{\lambda rX_4R}{a(m-wa)X}\,,\nn\\
\td{P}&=&P'+\frac{\lambda x(m-waY)P}{(m-wa)Y}\,,\nn\\
X_2&=&\lambda^2r^2+a^2(m-wa)^2\,,\nn\\
X_3&=&\lambda^2r^2-a^2(m-wa)^2\,,\nn\\
X_4&=&am-(r^2+a^2)w\,,\nn\\
Y_2&=&wa(m-wa)Y+\lambda (m-waY)^2\,,\nn\\
H_2&=&r^2+a^2(2-x^2)\,.\eea
%%%
The two functions, $R=R(r)$ and $P=P(x)\,$, satisfy
%%%
\bea R''&=&\frac{[\lambda^2+wa(m-wa)]XX_3-\lambda X_2X_4^2}{\lambda X^2X_2}R\nn\\
&&-\frac{(r^2-a^2)X_2-XX_3}{rXX_2}R'\,,\nn\\
P''&=&\frac{Y_2^2-\lambda^2Y[2wax^2(m+waY)+Y_2]}{\lambda Y^2Y_2}P\nn\\
&&+\frac{2\lambda x(m^2-w^2a^2Y^2)}{YY_2}P'\,.\eea
%%%
In this result, the separation constant has the clear meaning of being the eigenvalue of the operator $\cK_4$, though its
physical nature requires further investigation.

\section{Separating the linearized Einstein equations in Kerr}

In \ac{GR}, the vacuum equations governing the metric perturbations over the Kerr background (\ref{metric.kerr}) can be obtained from linearizing the Ricci tensor,
%%%
\bea (\wtd\Box h)_{\mu\nu}&\equiv&\frac12\nabla^\rho(\nabla_\mu h_{\nu\rho}+\nabla_\nu h_{\mu\rho}-\nabla_\rho h_{\mu\nu})-\frac12\nabla_\mu\nabla_\nu h\nn\\
&=&0\,,\label{eq.huv}\eea
%%%
where $h_{\mu\nu}\equiv\delta g_{\mu\nu}\,$.

Limited to the combinations that are both quadratic in the derivatives and in the Killing-Yano tensor, the following four operators can be found,
%%%
\bea(\cK_1h)_{\mu\nu}&=&\frac12(k^{\alpha\rho} k^\beta_{~\nu} -k^{\beta\rho} k^\alpha_{~\nu}
-k^{\alpha\beta} k^\rho_{~\nu}) \nabla_\alpha\nabla_\beta h_{\rho\mu}\nn\\
&&-\frac12 k^{\rho\alpha} k_{\rho\nu}[\nabla^\beta\,,\,\nabla_\mu]h_{\alpha\beta}
+\frac34k^{\alpha\beta}k^\rho_{~\nu}\nabla_\alpha\nabla_\beta h_{\rho\mu}\nn\\
&&+k_{\beta\rho} h_{\alpha\mu}\nabla_\nu\nabla^\rho k^{\alpha\beta}\nn\\
&& +\frac16 h_{\mu\nu} \nabla_\rho k_{\alpha\beta} \nabla^\rho k^{\alpha\beta}
+\frac14 h k^\beta_{~\nu} \nabla_\alpha \nabla^\alpha k_{\beta\mu}\,,\nn\\
(\cK_2h)_{\mu\nu}&=&(k_{\rho\nu}\nabla^\alpha k^\rho_{~\mu}+k_{\rho\mu}\nabla^\alpha k^\rho_{~\nu}) \nabla^\beta\tdh_{\alpha\beta}\nn\\
&&-k^{\rho\alpha}(k_{\rho\nu}\nabla_\mu \nabla^\beta \tdh_{\alpha\beta} +k_{\rho\mu}\nabla_\nu\nabla^\beta\tdh_{\alpha\beta})\,,\nn\\
(\cK_3h)_{\mu\nu}&=&\nabla_\mu(K\nabla^\rho\tdh_{\rho\nu}) +\nabla_\nu(K\nabla^\rho \tdh_{\rho\mu})\,,\nn\\
(\cK_4h)_{\mu\nu}&=&(k^{\alpha\rho} k^\beta_{~\nu} -k^{\beta\rho} k^\alpha_{~\nu})\nabla_\alpha\nabla_\beta h_{\rho\mu}\nn\\
&&+(k^{\alpha\rho} k^\beta_{~\mu} -k^{\beta\rho} k^\alpha_{~\mu})\nabla_\alpha\nabla_\beta h_{\rho\nu}\nn\\
&&-k^{\beta\rho}(k_{\alpha\rho}\nabla_\beta\nabla^\alpha h_{\mu\nu}+\nabla_\beta k_{\alpha\rho} \nabla^\alpha h_{\mu\nu})\nn\\
&&+\nabla^\alpha k^{\beta\rho}(h_{\alpha\mu}\nabla_\nu k_{\beta\rho}+ h_{\alpha\nu}\nabla_\mu k_{\beta\rho})\nn\\
&&-h^{\alpha\beta}(\nabla_\mu k_{\alpha\rho} \nabla_\nu k_\beta^{~\rho}
+\nabla_\nu k_{\alpha\rho}\nabla_\mu k_\beta^{~\rho})\nn\\
&&+2(J_\nu^{~\alpha\beta}-J^{\alpha~\beta}_{~\nu})\nabla_\beta h_{\alpha\mu}\nn\\
&&+2(J_\mu^{~\alpha\beta}-J^{\alpha~\beta}_{~\mu})\nabla_\beta h_{\alpha\nu}\,,\eea
%%%
where $\tdh_{\alpha\beta}=h_{\alpha\beta} -\frac12 g_{\alpha\beta}h\,$, $h=g^{\mu\nu}h_{\mu\nu}\,$, and $J^\alpha_{~\mu\nu}=k_{\rho\mu}\nabla^\alpha k^\rho_{~\nu} -k_{\rho\nu}\nabla^\alpha k^\rho_{~\mu}\,$.
Like in the vector case, the operators vary in their properties: $(\cK_1h)_{\mu\nu}\equiv h_{\mu\nu}$ and so is trivial; both $\cK_2$ and $\cK_3$ vanish in the de Donder gauge,
%%%
\bea\nabla^\alpha\tdh_{\alpha\beta}=\nabla^\alpha h_{\alpha\beta}-\frac12\pd_\beta h=0\,.\label{gauge.deDonder}\eea
%%%
Treating the Weyl scalars as operators,
%%%
\bea \psi_0=\psi_0[h_{\mu\nu}]\,,\quad \psi_4=\psi_4[h_{\mu\nu}]\,,\label{eom.weyle.scalar}\eea
%%%
where the detailed expressions for $\psi_0[h_{\mu\nu}]$ and $\psi_4[h_{\mu\nu}]$ can be found in \cite{Chrzanowski:1975wv}, one can also find that
%%%
\bea \psi_0\cK_i=\psi_4\cK_i=0\,,\quad i=2,3\,,\eea
%%%
for all solutions to (\ref{eq.huv}).
$\cK_4$ has none of these problems. In view of these properties, $\cK_4$ is used to do the separation of variables.

Before proceeding, note $(\cK_4h)_{\mu\nu}$ is not gauge invariant and $\cK_4$ generates new gauge transformations from existing ones,
%%%
\bea \cK_4\cL_\xi g_{\mu\nu}=\cL_{\xi'}g_{\mu\nu}\,,\label{def.Kxi1}\eea
%%%
where
%%%
\bea \xi'_\mu&=&\nabla^\alpha k^{\beta\gamma}(\xi_\alpha\nabla_\mu k_{\beta\gamma}
+2k_{\mu\alpha}\nabla_\beta\xi_\gamma -k_{\alpha\gamma}\nabla_\beta\xi_\mu)\nn\\
&&-K^{\alpha\beta}(\nabla_\alpha\nabla_\beta\xi_\mu -\nabla_\alpha\nabla_\mu\xi_\beta +\nabla_\mu\nabla_\alpha\xi_\beta)\nn\\
&&-2(\nabla_\alpha\xi_\beta-2\nabla_\beta\xi_\alpha)k^{\beta\gamma}\nabla^\alpha k_{\mu\gamma}\nn\\
&&+\xi^\alpha k^{\beta\gamma}\nabla_\alpha\nabla_\mu k_{\beta\gamma}\,.\label{def.Kxi2}\eea
%%%

In the following, I shall focus on the mode function $h^{(\lambda wm)}_{\mu\nu}$ defined in \eqref{decomp} and shall drop its superscripts $(\lambda wm)$.
The goal is to look for simultaneous solutions to (\ref{eq.huv}), (\ref{gauge.deDonder}) and the eigen equation,
%%%
\bea(\cK_4h)_{\mu\nu}=\lambda h_{\mu\nu}\,,\label{eom.eigen.guv}\eea
%%%
where $\lambda$ is a constant.
In an earlier effort (see the first version of this work at \cite{Mei:2023phov1}), a solution has been constructed but was soon found to be purely gauge.\footnote{The author thanks Nicola Franchini and Sam Dolan for helpful communications on this solution.}
It can be generated with
%%%
\bea \xi_\mu dx^\mu=d\Big[e^{i(wt-m\phi)}R(r)P(x)\Big]\,,\eea
%%%
where $R(r)$ and $P(x)$ are governed by (\ref{eom.master}) with $s=0$.
This has led to concerns that it might be too restrictive to impose (\ref{eq.huv}), (\ref{gauge.deDonder}) and (\ref{eom.eigen.guv}) simultaneously.
Fortunately, this concern has turned out to be unnecessary.

For the set of equations, (\ref{eq.huv}), (\ref{gauge.deDonder}) and (\ref{eom.eigen.guv}), if one expands $h_{\mu\nu}$ in the limit $x\rightarrow0$, then the coefficients at the $\cO(x^{n})$ order, $\forall\;n\geq2\,$, can always be solved {\it algebraically} in terms of those at the $\cO(x^{n-1})$ order and lower.
As a result, one only needs to deal with the differential equations for the coefficients at the lowest $\cO(x^{0})$ and $\cO(x^{1})$ orders.
This nice feature still holds if one chooses to expand in the limit $r\rightarrow0$ instead.
Series solutions found in such expansions indicate that one may consider the following ansatz:
%%%
\bea&&h_{\mu\nu}(t,r,x,\phi)=f_{\mu\nu}(r,x)e^{i(wt-m\phi)}\,,\nn\\
&&f_{\mu\nu}=\frac{f_{\mu\nu}^{(a)}(r,x)P(x) +f_{\mu\nu}^{(b)}(r,x)P'(x)}{H Y B_\mu B_\nu}\,,\nn\\
&&f_{\mu\nu}^{(a)}(r,x)=\sum_{q=1}^{12}f_{\mu\nu,q}^{(a)}(r)x^q\,,\nn\\
&&f_{\mu\nu}^{(b)}(r,x)=\sum_{q=1}^{12}f_{\mu\nu,q}^{(b)}(r)x^q\,,\nn\\
&&f_{\mu\nu,q}^{(a)}\,,\,f_{\mu\nu,q}^{(b)}\sim(\cdots)R(r)+(\cdots)R'(r)\,,\label{sol.guv}\eea
%%%
where $B_\mu=\{H,1,Y,H\}\,$, and $R(r)$ and $P(x)$ are governed by (\ref{eom.master}) with $s=\pm2$ and $\td\lambda=\lambda-s(s+1)\,$.
The relation for the separation constant $\lambda$ indicates that it is likely a characterization of the angular momentum, but further study is needed to precisely determine its nature.
Note the ansatz only requires $f_{\mu\nu}^{(a)}$ and $f_{\mu\nu}^{(b)}$ to be expanded to the 12th power of $x$.
If one tries to expand to a few higher powers, the corresponding coefficients will be found to vanish.

Due to the special feature of the equations explained above, all functions $f_{\mu\nu,q}^{(a)}$ and $f_{\mu\nu,q}^{(b)}$ with $q\geq2$ can be solved {\it algebraically} in terms of those with $q=0,1$ and their derivatives.
The only differential equations that need to be solved are those for $f_{\mu\nu,q}^{(a)}$ and $f_{\mu\nu,q}^{(b)}$ with $q=0,1$, resulting from plugging the details of those with $q\geq2$ back into equations (\ref{eq.huv}), (\ref{gauge.deDonder}) and (\ref{eom.eigen.guv}).
All calculations are done with the software {\sf Mathematica}.
Two independent solutions have been found.
The solutions are lengthy, and three accompanying files have been provided in the supplemental material, where {\sf s0.m} and {\sf s4.m} each contains the detail of an independent solution, and {\sf check.nb} contains some example code to check the results.

The solutions {\sf s0.m} and {\sf s4.m} appear to be complementary to each other.
For the solution {\sf s0.m}, the corresponding Weyl scalars are
%%%
\bea \psi_0=R(r)P(x)\,,\quad \psi_4=0\,,\label{prop.s0}\eea
%%%
where $R(r)$ and $P(x)$ are governed by (\ref{eom.master}) with $s=2\,$.
For that {\sf s4.m}, the corresponding Weyl scalars are
%%%
\bea \psi_0=0\,,\quad \psi_4=\frac{R(r)P(x)}{(r-iax)^4}\,,\label{prop.s4}\eea
%%%
where $R(r)$ and $P(x)$ are governed by (\ref{eom.master}) with $s=-2\,$.
More detailed studies of the properties of the solutions have been left to the future.

\section{Summary and discussion}

In this paper, I have tried to separate the linearized Einstein equations in Kerr directly by using the symmetry of the equations.
A symmetry operator that commutes with the wave operator has been constructed and explicitly presented.
The eigen value of the symmetry operator is expected to become the constant of separation appearing in the separated equations.
With the help of the symmetry operator, two explicit solutions of $f^{(\lambda wm)}_{\mu\nu}$ have been constructed showing how each metric component depends on the functions $R(r)$ and $P(x)$, which provides the necessary bridge between the linearized Einstein equations in Kerr and the two ordinary differential equations in \eqref{eom.master}.
A similar construction has also been carried out for separating the Klein-Gordon equation and the Maxwell equations in Kerr as examples.

Although the newly found symmetry operator has allowed us to directly separate the linearized Einstein equations without using metric reconstruction, which is something that has never been achieved before, the process still has difficulties.
It has been found that, even when the linearized equations (\ref{eq.huv}), the Lorenz gauge condition (\ref{gauge.deDonder}), and the eigen equation (\ref{eom.eigen.guv}) are imposed simultaneously, the equations still do not separate automatically.
In this paper, I have been forced to expand $f^{(\lambda wm)}_{\mu\nu}$ in the small $x$ limit first, and then tried to guess what the full structure of the solution should be.
For this reason, the generalities of the ansatz (\ref{sol.guv}) and the resultant solutions are not guaranteed, and more work is needed to either prove their general applicability or to make necessary extensions.

Despite this limitation, the method presented in this paper is still expected to provide a valuable alternative and a new starting point for future works that do not want to rely on metric reconstruction to separate the linearized Einstein equations in Kerr.
In addition, it is clear from (\ref{prop.s0}) and (\ref{prop.s4}) that the solution {\sf s0.m} ({\sf s4.m}) corresponds to the ingoing (outgoing) transverse radiations at large distances \cite{Szekeres:1965ux,Stephani:2003tm}.
So, although the generality of the two solutions are not proven, they are expected to describe (at least a subclass of) generic metric perturbations with both ingoing and outgoing radiations.
More detailed properties of the solutions are still being investigated.

\section*{Acknowledgments}

The author thanks Prof. Yun-Kau Lau for kindly hosting him at the Morningside Center of Mathematics, Chinese Academy of Sciences, for several weeks in 2014, when the author was first made aware of many interesting topics in GW research, including the problem investigated in this work.
The work has been supported in part by the National Key Research and Development Program of China (Grant No. 2023YFC2206700), the Natural Science Foundation of China (Grants No. 12261131504) and the Guangdong Major Project of Basic and Applied Basic Research (Grant No. 2019B030302001), and the Fundamental Research Funds for the Central Universities, Sun Yat-sen University.
\\

{\bf Conflict of interest} The authors declare that they have no conflict of interest.

%%%%%%%%%%%%%%%%%%%%%%%%%%%%%%%%%%%%%%%%%%%%%%%%%%%%%%%%%%%%%%%%
%\bibliographystyle{unsrt}
\bibliographystyle{apsrev4-1}
%%%%%%%%%%%%%%%%%%%%%%%%%%%%%%%%%%%%%%%%%%%%%%%%%%%%%%%%%%%%%%%%
\bibliography{ref}

\begin{thebibliography}{63}%
\makeatletter
\providecommand \@ifxundefined [1]{%
 \@ifx{#1\undefined}
}%
\providecommand \@ifnum [1]{%
 \ifnum #1\expandafter \@firstoftwo
 \else \expandafter \@secondoftwo
 \fi
}%
\providecommand \@ifx [1]{%
 \ifx #1\expandafter \@firstoftwo
 \else \expandafter \@secondoftwo
 \fi
}%
\providecommand \natexlab [1]{#1}%
\providecommand \enquote  [1]{``#1''}%
\providecommand \bibnamefont  [1]{#1}%
\providecommand \bibfnamefont [1]{#1}%
\providecommand \citenamefont [1]{#1}%
\providecommand \href@noop [0]{\@secondoftwo}%
\providecommand \href [0]{\begingroup \@sanitize@url \@href}%
\providecommand \@href[1]{\@@startlink{#1}\@@href}%
\providecommand \@@href[1]{\endgroup#1\@@endlink}%
\providecommand \@sanitize@url [0]{\catcode `\\12\catcode `\$12\catcode
  `\&12\catcode `\#12\catcode `\^12\catcode `\_12\catcode `\%12\relax}%
\providecommand \@@startlink[1]{}%
\providecommand \@@endlink[0]{}%
\providecommand \url  [0]{\begingroup\@sanitize@url \@url }%
\providecommand \@url [1]{\endgroup\@href {#1}{\urlprefix }}%
\providecommand \urlprefix  [0]{URL }%
\providecommand \Eprint [0]{\href }%
\providecommand \doibase [0]{http://dx.doi.org/}%
\providecommand \selectlanguage [0]{\@gobble}%
\providecommand \bibinfo  [0]{\@secondoftwo}%
\providecommand \bibfield  [0]{\@secondoftwo}%
\providecommand \translation [1]{[#1]}%
\providecommand \BibitemOpen [0]{}%
\providecommand \bibitemStop [0]{}%
\providecommand \bibitemNoStop [0]{.\EOS\space}%
\providecommand \EOS [0]{\spacefactor3000\relax}%
\providecommand \BibitemShut  [1]{\csname bibitem#1\endcsname}%
\let\auto@bib@innerbib\@empty
%</preamble>
\bibitem [{\citenamefont {Kormendy}\ and\ \citenamefont
  {Richstone}(1995)}]{Kormendy1995}%
  \BibitemOpen
  \bibfield  {author} {\bibinfo {author} {\bibfnamefont {J.}~\bibnamefont
  {Kormendy}}\ and\ \bibinfo {author} {\bibfnamefont {D.}~\bibnamefont
  {Richstone}},\ }\href {\doibase 10.1146/annurev.aa.33.090195.003053}
  {\bibfield  {journal} {\bibinfo  {journal} {Annual Review of Astronomy and
  Astrophysics}\ }\textbf {\bibinfo {volume} {33}},\ \bibinfo {pages} {581}
  (\bibinfo {year} {1995})},\ \Eprint
  {http://arxiv.org/abs/https://doi.org/10.1146/annurev.aa.33.090195.003053}
  {https://doi.org/10.1146/annurev.aa.33.090195.003053} \BibitemShut {NoStop}%
\bibitem [{\citenamefont {Elbert}\ \emph {et~al.}(2018)\citenamefont {Elbert},
  \citenamefont {Bullock},\ and\ \citenamefont {Kaplinghat}}]{Elbert:2017sbr}%
  \BibitemOpen
  \bibfield  {author} {\bibinfo {author} {\bibfnamefont {O.~D.}\ \bibnamefont
  {Elbert}}, \bibinfo {author} {\bibfnamefont {J.~S.}\ \bibnamefont {Bullock}},
  \ and\ \bibinfo {author} {\bibfnamefont {M.}~\bibnamefont {Kaplinghat}},\
  }\href {\doibase 10.1093/mnras/stx1959} {\bibfield  {journal} {\bibinfo
  {journal} {Mon. Not. Roy. Astron. Soc.}\ }\textbf {\bibinfo {volume} {473}},\
  \bibinfo {pages} {1186} (\bibinfo {year} {2018})},\ \Eprint
  {http://arxiv.org/abs/1703.02551} {arXiv:1703.02551 [astro-ph.GA]}
  \BibitemShut {NoStop}%
\bibitem [{\citenamefont {Gibbons}(1975)}]{Gibbons:1975kk}%
  \BibitemOpen
  \bibfield  {author} {\bibinfo {author} {\bibfnamefont {G.~W.}\ \bibnamefont
  {Gibbons}},\ }\href {\doibase 10.1007/BF01609829} {\bibfield  {journal}
  {\bibinfo  {journal} {Commun. Math. Phys.}\ }\textbf {\bibinfo {volume}
  {44}},\ \bibinfo {pages} {245} (\bibinfo {year} {1975})}\BibitemShut
  {NoStop}%
\bibitem [{\citenamefont {Ghosh}\ and\ \citenamefont
  {Afrin}(2023)}]{Ghosh:2022kit}%
  \BibitemOpen
  \bibfield  {author} {\bibinfo {author} {\bibfnamefont {S.~G.}\ \bibnamefont
  {Ghosh}}\ and\ \bibinfo {author} {\bibfnamefont {M.}~\bibnamefont {Afrin}},\
  }\href {\doibase 10.3847/1538-4357/acb695} {\bibfield  {journal} {\bibinfo
  {journal} {Astrophys. J.}\ }\textbf {\bibinfo {volume} {944}},\ \bibinfo
  {pages} {174} (\bibinfo {year} {2023})},\ \Eprint
  {http://arxiv.org/abs/2206.02488} {arXiv:2206.02488 [gr-qc]} \BibitemShut
  {NoStop}%
\bibitem [{\citenamefont {Carter}(1971)}]{Carter:1971zc}%
  \BibitemOpen
  \bibfield  {author} {\bibinfo {author} {\bibfnamefont {B.}~\bibnamefont
  {Carter}},\ }\href {\doibase 10.1103/PhysRevLett.26.331} {\bibfield
  {journal} {\bibinfo  {journal} {Phys. Rev. Lett.}\ }\textbf {\bibinfo
  {volume} {26}},\ \bibinfo {pages} {331} (\bibinfo {year} {1971})}\BibitemShut
  {NoStop}%
\bibitem [{\citenamefont {Bekenstein}(1996)}]{Bekenstein:1996pn}%
  \BibitemOpen
  \bibfield  {author} {\bibinfo {author} {\bibfnamefont {J.~D.}\ \bibnamefont
  {Bekenstein}},\ }in\ \href@noop {} {\emph {\bibinfo {booktitle} {{2nd
  International Sakharov Conference on Physics}}}}\ (\bibinfo {year} {1996})\
  pp.\ \bibinfo {pages} {216--219},\ \Eprint
  {http://arxiv.org/abs/gr-qc/9605059} {arXiv:gr-qc/9605059} \BibitemShut
  {NoStop}%
\bibitem [{\citenamefont {Chrusciel}\ \emph {et~al.}(2012)\citenamefont
  {Chrusciel}, \citenamefont {Lopes~Costa},\ and\ \citenamefont
  {Heusler}}]{Chrusciel:2012jk}%
  \BibitemOpen
  \bibfield  {author} {\bibinfo {author} {\bibfnamefont {P.~T.}\ \bibnamefont
  {Chrusciel}}, \bibinfo {author} {\bibfnamefont {J.}~\bibnamefont
  {Lopes~Costa}}, \ and\ \bibinfo {author} {\bibfnamefont {M.}~\bibnamefont
  {Heusler}},\ }\href {\doibase 10.12942/lrr-2012-7} {\bibfield  {journal}
  {\bibinfo  {journal} {Living Rev. Rel.}\ }\textbf {\bibinfo {volume} {15}},\
  \bibinfo {pages} {7} (\bibinfo {year} {2012})},\ \Eprint
  {http://arxiv.org/abs/1205.6112} {arXiv:1205.6112 [gr-qc]} \BibitemShut
  {NoStop}%
\bibitem [{\citenamefont {Cardoso}\ and\ \citenamefont
  {Gualtieri}(2016)}]{Cardoso:2016ryw}%
  \BibitemOpen
  \bibfield  {author} {\bibinfo {author} {\bibfnamefont {V.}~\bibnamefont
  {Cardoso}}\ and\ \bibinfo {author} {\bibfnamefont {L.}~\bibnamefont
  {Gualtieri}},\ }\href {\doibase 10.1088/0264-9381/33/17/174001} {\bibfield
  {journal} {\bibinfo  {journal} {Class. Quant. Grav.}\ }\textbf {\bibinfo
  {volume} {33}},\ \bibinfo {pages} {174001} (\bibinfo {year} {2016})},\
  \Eprint {http://arxiv.org/abs/1607.03133} {arXiv:1607.03133 [gr-qc]}
  \BibitemShut {NoStop}%
\bibitem [{\citenamefont {Kerr}(1963)}]{Kerr:1963ud}%
  \BibitemOpen
  \bibfield  {author} {\bibinfo {author} {\bibfnamefont {R.~P.}\ \bibnamefont
  {Kerr}},\ }\href {\doibase 10.1103/PhysRevLett.11.237} {\bibfield  {journal}
  {\bibinfo  {journal} {Phys. Rev. Lett.}\ }\textbf {\bibinfo {volume} {11}},\
  \bibinfo {pages} {237} (\bibinfo {year} {1963})}\BibitemShut {NoStop}%
\bibitem [{\citenamefont {Barack}\ \emph {et~al.}(2019)\citenamefont {Barack}
  \emph {et~al.}}]{Barack:2018yly}%
  \BibitemOpen
  \bibfield  {author} {\bibinfo {author} {\bibfnamefont {L.}~\bibnamefont
  {Barack}} \emph {et~al.},\ }\href {\doibase 10.1088/1361-6382/ab0587}
  {\bibfield  {journal} {\bibinfo  {journal} {Class. Quant. Grav.}\ }\textbf
  {\bibinfo {volume} {36}},\ \bibinfo {pages} {143001} (\bibinfo {year}
  {2019})},\ \Eprint {http://arxiv.org/abs/1806.05195} {arXiv:1806.05195
  [gr-qc]} \BibitemShut {NoStop}%
\bibitem [{\citenamefont {Luo}\ \emph {et~al.}(2025)\citenamefont {Luo} \emph
  {et~al.}}]{Luo:2025ewp}%
  \BibitemOpen
  \bibfield  {author} {\bibinfo {author} {\bibfnamefont {J.}~\bibnamefont
  {Luo}} \emph {et~al.},\ }\href@noop {} {\  (\bibinfo {year} {2025})},\
  \Eprint {http://arxiv.org/abs/2502.20138} {arXiv:2502.20138 [gr-qc]}
  \BibitemShut {NoStop}%
\bibitem [{\citenamefont {Berti}\ \emph {et~al.}(2025)\citenamefont {Berti}
  \emph {et~al.}}]{Berti:2025hly}%
  \BibitemOpen
  \bibfield  {author} {\bibinfo {author} {\bibfnamefont {E.}~\bibnamefont
  {Berti}} \emph {et~al.},\ }\href@noop {} {\  (\bibinfo {year} {2025})},\
  \Eprint {http://arxiv.org/abs/2505.23895} {arXiv:2505.23895 [gr-qc]}
  \BibitemShut {NoStop}%
\bibitem [{\citenamefont {Poisson}\ \emph {et~al.}(2011)\citenamefont
  {Poisson}, \citenamefont {Pound},\ and\ \citenamefont
  {Vega}}]{Poisson:2011nh}%
  \BibitemOpen
  \bibfield  {author} {\bibinfo {author} {\bibfnamefont {E.}~\bibnamefont
  {Poisson}}, \bibinfo {author} {\bibfnamefont {A.}~\bibnamefont {Pound}}, \
  and\ \bibinfo {author} {\bibfnamefont {I.}~\bibnamefont {Vega}},\ }\href
  {\doibase 10.12942/lrr-2011-7} {\bibfield  {journal} {\bibinfo  {journal}
  {Living Rev. Rel.}\ }\textbf {\bibinfo {volume} {14}},\ \bibinfo {pages} {7}
  (\bibinfo {year} {2011})},\ \Eprint {http://arxiv.org/abs/1102.0529}
  {arXiv:1102.0529 [gr-qc]} \BibitemShut {NoStop}%
\bibitem [{\citenamefont {Barack}\ and\ \citenamefont
  {Pound}(2019)}]{Barack:2018yvs}%
  \BibitemOpen
  \bibfield  {author} {\bibinfo {author} {\bibfnamefont {L.}~\bibnamefont
  {Barack}}\ and\ \bibinfo {author} {\bibfnamefont {A.}~\bibnamefont {Pound}},\
  }\href {\doibase 10.1088/1361-6633/aae552} {\bibfield  {journal} {\bibinfo
  {journal} {Rept. Prog. Phys.}\ }\textbf {\bibinfo {volume} {82}},\ \bibinfo
  {pages} {016904} (\bibinfo {year} {2019})},\ \Eprint
  {http://arxiv.org/abs/1805.10385} {arXiv:1805.10385 [gr-qc]} \BibitemShut
  {NoStop}%
\bibitem [{\citenamefont {Pound}\ and\ \citenamefont
  {Wardell}(2021)}]{Pound:2021qin}%
  \BibitemOpen
  \bibfield  {author} {\bibinfo {author} {\bibfnamefont {A.}~\bibnamefont
  {Pound}}\ and\ \bibinfo {author} {\bibfnamefont {B.}~\bibnamefont
  {Wardell}},\ }\href {\doibase 10.1007/978-981-15-4702-7\_38-1} {\  (\bibinfo
  {year} {2021}),\ 10.1007/978-981-15-4702-7\_38-1},\ \Eprint
  {http://arxiv.org/abs/2101.04592} {arXiv:2101.04592 [gr-qc]} \BibitemShut
  {NoStop}%
\bibitem [{\citenamefont {Cheung}\ \emph {et~al.}(2023)\citenamefont {Cheung}
  \emph {et~al.}}]{Cheung:2022rbm}%
  \BibitemOpen
  \bibfield  {author} {\bibinfo {author} {\bibfnamefont {M.~H.-Y.}\
  \bibnamefont {Cheung}} \emph {et~al.},\ }\href {\doibase
  10.1103/PhysRevLett.130.081401} {\bibfield  {journal} {\bibinfo  {journal}
  {Phys. Rev. Lett.}\ }\textbf {\bibinfo {volume} {130}},\ \bibinfo {pages}
  {081401} (\bibinfo {year} {2023})},\ \Eprint
  {http://arxiv.org/abs/2208.07374} {arXiv:2208.07374 [gr-qc]} \BibitemShut
  {NoStop}%
\bibitem [{\citenamefont {Mitman}\ \emph {et~al.}(2023)\citenamefont {Mitman}
  \emph {et~al.}}]{Mitman:2022qdl}%
  \BibitemOpen
  \bibfield  {author} {\bibinfo {author} {\bibfnamefont {K.}~\bibnamefont
  {Mitman}} \emph {et~al.},\ }\href {\doibase 10.1103/PhysRevLett.130.081402}
  {\bibfield  {journal} {\bibinfo  {journal} {Phys. Rev. Lett.}\ }\textbf
  {\bibinfo {volume} {130}},\ \bibinfo {pages} {081402} (\bibinfo {year}
  {2023})},\ \Eprint {http://arxiv.org/abs/2208.07380} {arXiv:2208.07380
  [gr-qc]} \BibitemShut {NoStop}%
\bibitem [{\citenamefont {Qiu}\ \emph {et~al.}(2024)\citenamefont {Qiu},
  \citenamefont {Forteza},\ and\ \citenamefont {Mourier}}]{Qiu:2023lwo}%
  \BibitemOpen
  \bibfield  {author} {\bibinfo {author} {\bibfnamefont {Y.}~\bibnamefont
  {Qiu}}, \bibinfo {author} {\bibfnamefont {X.~J.}\ \bibnamefont {Forteza}}, \
  and\ \bibinfo {author} {\bibfnamefont {P.}~\bibnamefont {Mourier}},\ }\href
  {\doibase 10.1103/PhysRevD.109.064075} {\bibfield  {journal} {\bibinfo
  {journal} {Phys. Rev. D}\ }\textbf {\bibinfo {volume} {109}},\ \bibinfo
  {pages} {064075} (\bibinfo {year} {2024})},\ \Eprint
  {http://arxiv.org/abs/2312.15904} {arXiv:2312.15904 [gr-qc]} \BibitemShut
  {NoStop}%
\bibitem [{\citenamefont {Shi}\ \emph {et~al.}(2024)\citenamefont {Shi},
  \citenamefont {Zhang},\ and\ \citenamefont {Mei}}]{Shi:2024ttu}%
  \BibitemOpen
  \bibfield  {author} {\bibinfo {author} {\bibfnamefont {C.}~\bibnamefont
  {Shi}}, \bibinfo {author} {\bibfnamefont {Q.}~\bibnamefont {Zhang}}, \ and\
  \bibinfo {author} {\bibfnamefont {J.}~\bibnamefont {Mei}},\ }\href {\doibase
  10.1103/PhysRevD.110.124007} {\bibfield  {journal} {\bibinfo  {journal}
  {Phys. Rev. D}\ }\textbf {\bibinfo {volume} {110}},\ \bibinfo {pages}
  {124007} (\bibinfo {year} {2024})},\ \Eprint
  {http://arxiv.org/abs/2407.13110} {arXiv:2407.13110 [gr-qc]} \BibitemShut
  {NoStop}%
\bibitem [{\citenamefont {Yi}\ \emph {et~al.}(2024)\citenamefont {Yi},
  \citenamefont {Kuntz}, \citenamefont {Barausse}, \citenamefont {Berti},
  \citenamefont {Cheung}, \citenamefont {Kritos},\ and\ \citenamefont
  {Maselli}}]{Yi:2024elj}%
  \BibitemOpen
  \bibfield  {author} {\bibinfo {author} {\bibfnamefont {S.}~\bibnamefont
  {Yi}}, \bibinfo {author} {\bibfnamefont {A.}~\bibnamefont {Kuntz}}, \bibinfo
  {author} {\bibfnamefont {E.}~\bibnamefont {Barausse}}, \bibinfo {author}
  {\bibfnamefont {E.}~\bibnamefont {Berti}}, \bibinfo {author} {\bibfnamefont
  {M.~H.-Y.}\ \bibnamefont {Cheung}}, \bibinfo {author} {\bibfnamefont
  {K.}~\bibnamefont {Kritos}}, \ and\ \bibinfo {author} {\bibfnamefont
  {A.}~\bibnamefont {Maselli}},\ }\href {\doibase 10.1103/PhysRevD.109.124029}
  {\bibfield  {journal} {\bibinfo  {journal} {Phys. Rev. D}\ }\textbf {\bibinfo
  {volume} {109}},\ \bibinfo {pages} {124029} (\bibinfo {year} {2024})},\
  \Eprint {http://arxiv.org/abs/2403.09767} {arXiv:2403.09767 [gr-qc]}
  \BibitemShut {NoStop}%
\bibitem [{\citenamefont {Khera}\ \emph {et~al.}(2025)\citenamefont {Khera},
  \citenamefont {Ma},\ and\ \citenamefont {Yang}}]{Khera:2024bjs}%
  \BibitemOpen
  \bibfield  {author} {\bibinfo {author} {\bibfnamefont {N.}~\bibnamefont
  {Khera}}, \bibinfo {author} {\bibfnamefont {S.}~\bibnamefont {Ma}}, \ and\
  \bibinfo {author} {\bibfnamefont {H.}~\bibnamefont {Yang}},\ }\href {\doibase
  10.1103/PhysRevLett.134.211404} {\bibfield  {journal} {\bibinfo  {journal}
  {Phys. Rev. Lett.}\ }\textbf {\bibinfo {volume} {134}},\ \bibinfo {pages}
  {211404} (\bibinfo {year} {2025})},\ \Eprint
  {http://arxiv.org/abs/2410.14529} {arXiv:2410.14529 [gr-qc]} \BibitemShut
  {NoStop}%
\bibitem [{\citenamefont {Lagos}\ \emph {et~al.}(2025)\citenamefont {Lagos},
  \citenamefont {Andrade}, \citenamefont {Rafecas-Ventosa},\ and\ \citenamefont
  {Hui}}]{Lagos:2024ekd}%
  \BibitemOpen
  \bibfield  {author} {\bibinfo {author} {\bibfnamefont {M.}~\bibnamefont
  {Lagos}}, \bibinfo {author} {\bibfnamefont {T.}~\bibnamefont {Andrade}},
  \bibinfo {author} {\bibfnamefont {J.}~\bibnamefont {Rafecas-Ventosa}}, \ and\
  \bibinfo {author} {\bibfnamefont {L.}~\bibnamefont {Hui}},\ }\href {\doibase
  10.1103/PhysRevD.111.024018} {\bibfield  {journal} {\bibinfo  {journal}
  {Phys. Rev. D}\ }\textbf {\bibinfo {volume} {111}},\ \bibinfo {pages}
  {024018} (\bibinfo {year} {2025})},\ \Eprint
  {http://arxiv.org/abs/2411.02264} {arXiv:2411.02264 [gr-qc]} \BibitemShut
  {NoStop}%
\bibitem [{\citenamefont {Arun}\ \emph {et~al.}(2022)\citenamefont {Arun} \emph
  {et~al.}}]{LISA:2022kgy}%
  \BibitemOpen
  \bibfield  {author} {\bibinfo {author} {\bibfnamefont {K.~G.}\ \bibnamefont
  {Arun}} \emph {et~al.} (\bibinfo {collaboration} {LISA}),\ }\href {\doibase
  10.1007/s41114-022-00036-9} {\bibfield  {journal} {\bibinfo  {journal}
  {Living Rev. Rel.}\ }\textbf {\bibinfo {volume} {25}},\ \bibinfo {pages} {4}
  (\bibinfo {year} {2022})},\ \Eprint {http://arxiv.org/abs/2205.01597}
  {arXiv:2205.01597 [gr-qc]} \BibitemShut {NoStop}%
\bibitem [{\citenamefont {Kalogera}\ \emph {et~al.}(2021)\citenamefont
  {Kalogera} \emph {et~al.}}]{Kalogera:2021bya}%
  \BibitemOpen
  \bibfield  {author} {\bibinfo {author} {\bibfnamefont {V.}~\bibnamefont
  {Kalogera}} \emph {et~al.},\ }\href@noop {} {\  (\bibinfo {year} {2021})},\
  \Eprint {http://arxiv.org/abs/2111.06990} {arXiv:2111.06990 [gr-qc]}
  \BibitemShut {NoStop}%
\bibitem [{\citenamefont {Reitze}\ \emph {et~al.}(2019)\citenamefont {Reitze}
  \emph {et~al.}}]{Reitze:2019iox}%
  \BibitemOpen
  \bibfield  {author} {\bibinfo {author} {\bibfnamefont {D.}~\bibnamefont
  {Reitze}} \emph {et~al.},\ }\href@noop {} {\bibfield  {journal} {\bibinfo
  {journal} {Bull. Am. Astron. Soc.}\ }\textbf {\bibinfo {volume} {51}},\
  \bibinfo {pages} {035} (\bibinfo {year} {2019})},\ \Eprint
  {http://arxiv.org/abs/1907.04833} {arXiv:1907.04833 [astro-ph.IM]}
  \BibitemShut {NoStop}%
\bibitem [{\citenamefont {Punturo}\ \emph {et~al.}(2010)\citenamefont {Punturo}
  \emph {et~al.}}]{Punturo:2010zz}%
  \BibitemOpen
  \bibfield  {author} {\bibinfo {author} {\bibfnamefont {M.}~\bibnamefont
  {Punturo}} \emph {et~al.},\ }\href {\doibase 10.1088/0264-9381/27/19/194002}
  {\bibfield  {journal} {\bibinfo  {journal} {Class. Quant. Grav.}\ }\textbf
  {\bibinfo {volume} {27}},\ \bibinfo {pages} {194002} (\bibinfo {year}
  {2010})}\BibitemShut {NoStop}%
\bibitem [{\citenamefont {Teukolsky}(1972)}]{Teukolsky:1972my}%
  \BibitemOpen
  \bibfield  {author} {\bibinfo {author} {\bibfnamefont {S.~A.}\ \bibnamefont
  {Teukolsky}},\ }\href {\doibase 10.1103/PhysRevLett.29.1114} {\bibfield
  {journal} {\bibinfo  {journal} {Phys. Rev. Lett.}\ }\textbf {\bibinfo
  {volume} {29}},\ \bibinfo {pages} {1114} (\bibinfo {year}
  {1972})}\BibitemShut {NoStop}%
%%CITATION = PRLTA,29,1114;%%
\bibitem [{\citenamefont {Teukolsky}(1973)}]{Teukolsky:1973ha}%
  \BibitemOpen
  \bibfield  {author} {\bibinfo {author} {\bibfnamefont {S.~A.}\ \bibnamefont
  {Teukolsky}},\ }\href {\doibase 10.1086/152444} {\bibfield  {journal}
  {\bibinfo  {journal} {Astrophys. J.}\ }\textbf {\bibinfo {volume} {185}},\
  \bibinfo {pages} {635} (\bibinfo {year} {1973})}\BibitemShut {NoStop}%
\bibitem [{\citenamefont {Teukolsky}(2015)}]{Teukolsky:2014vca}%
  \BibitemOpen
  \bibfield  {author} {\bibinfo {author} {\bibfnamefont {S.~A.}\ \bibnamefont
  {Teukolsky}},\ }\href {\doibase 10.1088/0264-9381/32/12/124006} {\bibfield
  {journal} {\bibinfo  {journal} {Class. Quant. Grav.}\ }\textbf {\bibinfo
  {volume} {32}},\ \bibinfo {pages} {124006} (\bibinfo {year} {2015})},\
  \Eprint {http://arxiv.org/abs/1410.2130} {arXiv:1410.2130 [gr-qc]}
  \BibitemShut {NoStop}%
\bibitem [{\citenamefont {Chrzanowski}(1975)}]{Chrzanowski:1975wv}%
  \BibitemOpen
  \bibfield  {author} {\bibinfo {author} {\bibfnamefont {P.~L.}\ \bibnamefont
  {Chrzanowski}},\ }\href {\doibase 10.1103/PhysRevD.11.2042} {\bibfield
  {journal} {\bibinfo  {journal} {Phys. Rev.}\ }\textbf {\bibinfo {volume}
  {D11}},\ \bibinfo {pages} {2042} (\bibinfo {year} {1975})}\BibitemShut
  {NoStop}%
%%CITATION = PHRVA,D11,2042;%%
\bibitem [{\citenamefont {Kegeles}\ and\ \citenamefont
  {Cohen}(1979)}]{Kegeles:1979an}%
  \BibitemOpen
  \bibfield  {author} {\bibinfo {author} {\bibfnamefont {L.~S.}\ \bibnamefont
  {Kegeles}}\ and\ \bibinfo {author} {\bibfnamefont {J.~M.}\ \bibnamefont
  {Cohen}},\ }\href {\doibase 10.1103/PhysRevD.19.1641} {\bibfield  {journal}
  {\bibinfo  {journal} {Phys. Rev.}\ }\textbf {\bibinfo {volume} {D19}},\
  \bibinfo {pages} {1641} (\bibinfo {year} {1979})}\BibitemShut {NoStop}%
%%CITATION = PHRVA,D19,1641;%%
\bibitem [{\citenamefont {Wald}(1978)}]{Wald:1978vm}%
  \BibitemOpen
  \bibfield  {author} {\bibinfo {author} {\bibfnamefont {R.~M.}\ \bibnamefont
  {Wald}},\ }\href {\doibase 10.1103/PhysRevLett.41.203} {\bibfield  {journal}
  {\bibinfo  {journal} {Phys. Rev. Lett.}\ }\textbf {\bibinfo {volume} {41}},\
  \bibinfo {pages} {203} (\bibinfo {year} {1978})}\BibitemShut {NoStop}%
%%CITATION = PRLTA,41,203;%%
\bibitem [{\citenamefont {Stewart}(1979)}]{Stewart:1978tm}%
  \BibitemOpen
  \bibfield  {author} {\bibinfo {author} {\bibfnamefont {J.~M.}\ \bibnamefont
  {Stewart}},\ }\href {\doibase 10.1098/rspa.1979.0101} {\bibfield  {journal}
  {\bibinfo  {journal} {Proc. Roy. Soc. Lond.}\ }\textbf {\bibinfo {volume}
  {A367}},\ \bibinfo {pages} {527} (\bibinfo {year} {1979})}\BibitemShut
  {NoStop}%
%%CITATION = PRSLA,A367,527;%%
\bibitem [{\citenamefont {Dolan}\ \emph {et~al.}(2022)\citenamefont {Dolan},
  \citenamefont {Kavanagh},\ and\ \citenamefont {Wardell}}]{Dolan:2021ijg}%
  \BibitemOpen
  \bibfield  {author} {\bibinfo {author} {\bibfnamefont {S.~R.}\ \bibnamefont
  {Dolan}}, \bibinfo {author} {\bibfnamefont {C.}~\bibnamefont {Kavanagh}}, \
  and\ \bibinfo {author} {\bibfnamefont {B.}~\bibnamefont {Wardell}},\ }\href
  {\doibase 10.1103/PhysRevLett.128.151101} {\bibfield  {journal} {\bibinfo
  {journal} {Phys. Rev. Lett.}\ }\textbf {\bibinfo {volume} {128}},\ \bibinfo
  {pages} {151101} (\bibinfo {year} {2022})},\ \Eprint
  {http://arxiv.org/abs/2108.06344} {arXiv:2108.06344 [gr-qc]} \BibitemShut
  {NoStop}%
\bibitem [{\citenamefont {Lousto}\ and\ \citenamefont
  {Whiting}(2002)}]{Lousto:2002em}%
  \BibitemOpen
  \bibfield  {author} {\bibinfo {author} {\bibfnamefont {C.~O.}\ \bibnamefont
  {Lousto}}\ and\ \bibinfo {author} {\bibfnamefont {B.~F.}\ \bibnamefont
  {Whiting}},\ }\href {\doibase 10.1103/PhysRevD.66.024026} {\bibfield
  {journal} {\bibinfo  {journal} {Phys. Rev. D}\ }\textbf {\bibinfo {volume}
  {66}},\ \bibinfo {pages} {024026} (\bibinfo {year} {2002})},\ \Eprint
  {http://arxiv.org/abs/gr-qc/0203061} {arXiv:gr-qc/0203061} \BibitemShut
  {NoStop}%
\bibitem [{\citenamefont {Berens}\ \emph {et~al.}(2024)\citenamefont {Berens},
  \citenamefont {Gravely},\ and\ \citenamefont {Lupsasca}}]{Berens:2024czo}%
  \BibitemOpen
  \bibfield  {author} {\bibinfo {author} {\bibfnamefont {R.}~\bibnamefont
  {Berens}}, \bibinfo {author} {\bibfnamefont {T.}~\bibnamefont {Gravely}}, \
  and\ \bibinfo {author} {\bibfnamefont {A.}~\bibnamefont {Lupsasca}},\ }\href
  {\doibase 10.1088/1361-6382/ad6c9c} {\bibfield  {journal} {\bibinfo
  {journal} {Class. Quant. Grav.}\ }\textbf {\bibinfo {volume} {41}},\ \bibinfo
  {pages} {195004} (\bibinfo {year} {2024})},\ \Eprint
  {http://arxiv.org/abs/2403.20311} {arXiv:2403.20311 [gr-qc]} \BibitemShut
  {NoStop}%
\bibitem [{\citenamefont {Merlin}\ \emph {et~al.}(2016)\citenamefont {Merlin},
  \citenamefont {Ori}, \citenamefont {Barack}, \citenamefont {Pound},\ and\
  \citenamefont {van~de Meent}}]{Merlin:2016boc}%
  \BibitemOpen
  \bibfield  {author} {\bibinfo {author} {\bibfnamefont {C.}~\bibnamefont
  {Merlin}}, \bibinfo {author} {\bibfnamefont {A.}~\bibnamefont {Ori}},
  \bibinfo {author} {\bibfnamefont {L.}~\bibnamefont {Barack}}, \bibinfo
  {author} {\bibfnamefont {A.}~\bibnamefont {Pound}}, \ and\ \bibinfo {author}
  {\bibfnamefont {M.}~\bibnamefont {van~de Meent}},\ }\href {\doibase
  10.1103/PhysRevD.94.104066} {\bibfield  {journal} {\bibinfo  {journal} {Phys.
  Rev. D}\ }\textbf {\bibinfo {volume} {94}},\ \bibinfo {pages} {104066}
  (\bibinfo {year} {2016})},\ \Eprint {http://arxiv.org/abs/1609.01227}
  {arXiv:1609.01227 [gr-qc]} \BibitemShut {NoStop}%
\bibitem [{\citenamefont {Green}\ \emph {et~al.}(2020)\citenamefont {Green},
  \citenamefont {Hollands},\ and\ \citenamefont {Zimmerman}}]{Green:2019nam}%
  \BibitemOpen
  \bibfield  {author} {\bibinfo {author} {\bibfnamefont {S.~R.}\ \bibnamefont
  {Green}}, \bibinfo {author} {\bibfnamefont {S.}~\bibnamefont {Hollands}}, \
  and\ \bibinfo {author} {\bibfnamefont {P.}~\bibnamefont {Zimmerman}},\ }\href
  {\doibase 10.1088/1361-6382/ab7075} {\bibfield  {journal} {\bibinfo
  {journal} {Class. Quant. Grav.}\ }\textbf {\bibinfo {volume} {37}},\ \bibinfo
  {pages} {075001} (\bibinfo {year} {2020})},\ \Eprint
  {http://arxiv.org/abs/1908.09095} {arXiv:1908.09095 [gr-qc]} \BibitemShut
  {NoStop}%
\bibitem [{\citenamefont {Loutrel}\ \emph {et~al.}(2021)\citenamefont
  {Loutrel}, \citenamefont {Ripley}, \citenamefont {Giorgi},\ and\
  \citenamefont {Pretorius}}]{Loutrel:2020wbw}%
  \BibitemOpen
  \bibfield  {author} {\bibinfo {author} {\bibfnamefont {N.}~\bibnamefont
  {Loutrel}}, \bibinfo {author} {\bibfnamefont {J.~L.}\ \bibnamefont {Ripley}},
  \bibinfo {author} {\bibfnamefont {E.}~\bibnamefont {Giorgi}}, \ and\ \bibinfo
  {author} {\bibfnamefont {F.}~\bibnamefont {Pretorius}},\ }\href {\doibase
  10.1103/PhysRevD.103.104017} {\bibfield  {journal} {\bibinfo  {journal}
  {Phys. Rev. D}\ }\textbf {\bibinfo {volume} {103}},\ \bibinfo {pages}
  {104017} (\bibinfo {year} {2021})},\ \Eprint
  {http://arxiv.org/abs/2008.11770} {arXiv:2008.11770 [gr-qc]} \BibitemShut
  {NoStop}%
\bibitem [{\citenamefont {Toomani}\ \emph {et~al.}(2022)\citenamefont
  {Toomani}, \citenamefont {Zimmerman}, \citenamefont {Spiers}, \citenamefont
  {Hollands}, \citenamefont {Pound},\ and\ \citenamefont
  {Green}}]{Toomani:2021jlo}%
  \BibitemOpen
  \bibfield  {author} {\bibinfo {author} {\bibfnamefont {V.}~\bibnamefont
  {Toomani}}, \bibinfo {author} {\bibfnamefont {P.}~\bibnamefont {Zimmerman}},
  \bibinfo {author} {\bibfnamefont {A.}~\bibnamefont {Spiers}}, \bibinfo
  {author} {\bibfnamefont {S.}~\bibnamefont {Hollands}}, \bibinfo {author}
  {\bibfnamefont {A.}~\bibnamefont {Pound}}, \ and\ \bibinfo {author}
  {\bibfnamefont {S.~R.}\ \bibnamefont {Green}},\ }\href {\doibase
  10.1088/1361-6382/ac37a5} {\bibfield  {journal} {\bibinfo  {journal} {Class.
  Quant. Grav.}\ }\textbf {\bibinfo {volume} {39}},\ \bibinfo {pages} {015019}
  (\bibinfo {year} {2022})},\ \Eprint {http://arxiv.org/abs/2108.04273}
  {arXiv:2108.04273 [gr-qc]} \BibitemShut {NoStop}%
\bibitem [{\citenamefont {Aly}\ and\ \citenamefont
  {Stojkovic}(2023)}]{Aly:2023qzk}%
  \BibitemOpen
  \bibfield  {author} {\bibinfo {author} {\bibfnamefont {F.}~\bibnamefont
  {Aly}}\ and\ \bibinfo {author} {\bibfnamefont {D.}~\bibnamefont
  {Stojkovic}},\ }\href {\doibase 10.1088/1361-6382/ad0495} {\bibfield
  {journal} {\bibinfo  {journal} {Class. Quant. Grav.}\ }\textbf {\bibinfo
  {volume} {40}},\ \bibinfo {pages} {235010} (\bibinfo {year} {2023})},\
  \Eprint {http://arxiv.org/abs/2309.09474} {arXiv:2309.09474 [gr-qc]}
  \BibitemShut {NoStop}%
\bibitem [{\citenamefont {Wardell}\ \emph {et~al.}(2023)\citenamefont
  {Wardell}, \citenamefont {Pound}, \citenamefont {Warburton}, \citenamefont
  {Miller}, \citenamefont {Durkan},\ and\ \citenamefont
  {Le~Tiec}}]{Wardell:2021fyy}%
  \BibitemOpen
  \bibfield  {author} {\bibinfo {author} {\bibfnamefont {B.}~\bibnamefont
  {Wardell}}, \bibinfo {author} {\bibfnamefont {A.}~\bibnamefont {Pound}},
  \bibinfo {author} {\bibfnamefont {N.}~\bibnamefont {Warburton}}, \bibinfo
  {author} {\bibfnamefont {J.}~\bibnamefont {Miller}}, \bibinfo {author}
  {\bibfnamefont {L.}~\bibnamefont {Durkan}}, \ and\ \bibinfo {author}
  {\bibfnamefont {A.}~\bibnamefont {Le~Tiec}},\ }\href {\doibase
  10.1103/PhysRevLett.130.241402} {\bibfield  {journal} {\bibinfo  {journal}
  {Phys. Rev. Lett.}\ }\textbf {\bibinfo {volume} {130}},\ \bibinfo {pages}
  {241402} (\bibinfo {year} {2023})},\ \Eprint
  {http://arxiv.org/abs/2112.12265} {arXiv:2112.12265 [gr-qc]} \BibitemShut
  {NoStop}%
\bibitem [{\citenamefont {Ripley}\ \emph {et~al.}(2021)\citenamefont {Ripley},
  \citenamefont {Loutrel}, \citenamefont {Giorgi},\ and\ \citenamefont
  {Pretorius}}]{Ripley:2020xby}%
  \BibitemOpen
  \bibfield  {author} {\bibinfo {author} {\bibfnamefont {J.~L.}\ \bibnamefont
  {Ripley}}, \bibinfo {author} {\bibfnamefont {N.}~\bibnamefont {Loutrel}},
  \bibinfo {author} {\bibfnamefont {E.}~\bibnamefont {Giorgi}}, \ and\ \bibinfo
  {author} {\bibfnamefont {F.}~\bibnamefont {Pretorius}},\ }\href {\doibase
  10.1103/PhysRevD.103.104018} {\bibfield  {journal} {\bibinfo  {journal}
  {Phys. Rev. D}\ }\textbf {\bibinfo {volume} {103}},\ \bibinfo {pages}
  {104018} (\bibinfo {year} {2021})},\ \Eprint
  {http://arxiv.org/abs/2010.00162} {arXiv:2010.00162 [gr-qc]} \BibitemShut
  {NoStop}%
\bibitem [{\citenamefont {Dolan}\ \emph {et~al.}(2023)\citenamefont {Dolan},
  \citenamefont {Durkan}, \citenamefont {Kavanagh},\ and\ \citenamefont
  {Wardell}}]{Dolan:2023enf}%
  \BibitemOpen
  \bibfield  {author} {\bibinfo {author} {\bibfnamefont {S.~R.}\ \bibnamefont
  {Dolan}}, \bibinfo {author} {\bibfnamefont {L.}~\bibnamefont {Durkan}},
  \bibinfo {author} {\bibfnamefont {C.}~\bibnamefont {Kavanagh}}, \ and\
  \bibinfo {author} {\bibfnamefont {B.}~\bibnamefont {Wardell}},\ }\href@noop
  {} {\  (\bibinfo {year} {2023})},\ \Eprint {http://arxiv.org/abs/2306.16459}
  {arXiv:2306.16459 [gr-qc]} \BibitemShut {NoStop}%
\bibitem [{\citenamefont {Sberna}\ \emph {et~al.}(2022)\citenamefont {Sberna},
  \citenamefont {Bosch}, \citenamefont {East}, \citenamefont {Green},\ and\
  \citenamefont {Lehner}}]{Sberna:2021eui}%
  \BibitemOpen
  \bibfield  {author} {\bibinfo {author} {\bibfnamefont {L.}~\bibnamefont
  {Sberna}}, \bibinfo {author} {\bibfnamefont {P.}~\bibnamefont {Bosch}},
  \bibinfo {author} {\bibfnamefont {W.~E.}\ \bibnamefont {East}}, \bibinfo
  {author} {\bibfnamefont {S.~R.}\ \bibnamefont {Green}}, \ and\ \bibinfo
  {author} {\bibfnamefont {L.}~\bibnamefont {Lehner}},\ }\href {\doibase
  10.1103/PhysRevD.105.064046} {\bibfield  {journal} {\bibinfo  {journal}
  {Phys. Rev. D}\ }\textbf {\bibinfo {volume} {105}},\ \bibinfo {pages}
  {064046} (\bibinfo {year} {2022})},\ \Eprint
  {http://arxiv.org/abs/2112.11168} {arXiv:2112.11168 [gr-qc]} \BibitemShut
  {NoStop}%
\bibitem [{\citenamefont {Yang}\ \emph {et~al.}(2015)\citenamefont {Yang},
  \citenamefont {Zimmerman},\ and\ \citenamefont {Lehner}}]{Yang:2014tla}%
  \BibitemOpen
  \bibfield  {author} {\bibinfo {author} {\bibfnamefont {H.}~\bibnamefont
  {Yang}}, \bibinfo {author} {\bibfnamefont {A.}~\bibnamefont {Zimmerman}}, \
  and\ \bibinfo {author} {\bibfnamefont {L.}~\bibnamefont {Lehner}},\ }\href
  {\doibase 10.1103/PhysRevLett.114.081101} {\bibfield  {journal} {\bibinfo
  {journal} {Phys. Rev. Lett.}\ }\textbf {\bibinfo {volume} {114}},\ \bibinfo
  {pages} {081101} (\bibinfo {year} {2015})},\ \Eprint
  {http://arxiv.org/abs/1402.4859} {arXiv:1402.4859 [gr-qc]} \BibitemShut
  {NoStop}%
\bibitem [{\citenamefont {Green}\ \emph {et~al.}(2023)\citenamefont {Green},
  \citenamefont {Hollands}, \citenamefont {Sberna}, \citenamefont {Toomani},\
  and\ \citenamefont {Zimmerman}}]{Green:2022htq}%
  \BibitemOpen
  \bibfield  {author} {\bibinfo {author} {\bibfnamefont {S.~R.}\ \bibnamefont
  {Green}}, \bibinfo {author} {\bibfnamefont {S.}~\bibnamefont {Hollands}},
  \bibinfo {author} {\bibfnamefont {L.}~\bibnamefont {Sberna}}, \bibinfo
  {author} {\bibfnamefont {V.}~\bibnamefont {Toomani}}, \ and\ \bibinfo
  {author} {\bibfnamefont {P.}~\bibnamefont {Zimmerman}},\ }\href {\doibase
  10.1103/PhysRevD.107.064030} {\bibfield  {journal} {\bibinfo  {journal}
  {Phys. Rev. D}\ }\textbf {\bibinfo {volume} {107}},\ \bibinfo {pages}
  {064030} (\bibinfo {year} {2023})},\ \Eprint
  {http://arxiv.org/abs/2210.15935} {arXiv:2210.15935 [gr-qc]} \BibitemShut
  {NoStop}%
\bibitem [{\citenamefont {Spiers}(2024)}]{Spiers:2024src}%
  \BibitemOpen
  \bibfield  {author} {\bibinfo {author} {\bibfnamefont {A.}~\bibnamefont
  {Spiers}},\ }\href {\doibase 10.1103/PhysRevD.109.104059} {\bibfield
  {journal} {\bibinfo  {journal} {Phys. Rev. D}\ }\textbf {\bibinfo {volume}
  {109}},\ \bibinfo {pages} {104059} (\bibinfo {year} {2024})},\ \Eprint
  {http://arxiv.org/abs/2402.00604} {arXiv:2402.00604 [gr-qc]} \BibitemShut
  {NoStop}%
\bibitem [{\citenamefont {Chung}\ \emph {et~al.}(2024)\citenamefont {Chung},
  \citenamefont {Wagle},\ and\ \citenamefont {Yunes}}]{Chung:2023wkd}%
  \BibitemOpen
  \bibfield  {author} {\bibinfo {author} {\bibfnamefont {A.~K.-W.}\
  \bibnamefont {Chung}}, \bibinfo {author} {\bibfnamefont {P.}~\bibnamefont
  {Wagle}}, \ and\ \bibinfo {author} {\bibfnamefont {N.}~\bibnamefont
  {Yunes}},\ }\href {\doibase 10.1103/PhysRevD.109.044072} {\bibfield
  {journal} {\bibinfo  {journal} {Phys. Rev. D}\ }\textbf {\bibinfo {volume}
  {109}},\ \bibinfo {pages} {044072} (\bibinfo {year} {2024})},\ \Eprint
  {http://arxiv.org/abs/2312.08435} {arXiv:2312.08435 [gr-qc]} \BibitemShut
  {NoStop}%
\bibitem [{\citenamefont {Hollands}\ and\ \citenamefont
  {Toomani}(2024)}]{Hollands:2024iqp}%
  \BibitemOpen
  \bibfield  {author} {\bibinfo {author} {\bibfnamefont {S.}~\bibnamefont
  {Hollands}}\ and\ \bibinfo {author} {\bibfnamefont {V.}~\bibnamefont
  {Toomani}},\ }\href {\doibase 10.1088/1361-6382/ad87a1} {\  (\bibinfo {year}
  {2024}),\ 10.1088/1361-6382/ad87a1},\ \Eprint
  {http://arxiv.org/abs/2405.18604} {arXiv:2405.18604 [gr-qc]} \BibitemShut
  {NoStop}%
\bibitem [{\citenamefont {Chen}\ and\ \citenamefont
  {Stein}(2017)}]{Chen:2017ofv}%
  \BibitemOpen
  \bibfield  {author} {\bibinfo {author} {\bibfnamefont {B.}~\bibnamefont
  {Chen}}\ and\ \bibinfo {author} {\bibfnamefont {L.~C.}\ \bibnamefont
  {Stein}},\ }\href {\doibase 10.1103/PhysRevD.96.064017} {\bibfield  {journal}
  {\bibinfo  {journal} {Phys. Rev. D}\ }\textbf {\bibinfo {volume} {96}},\
  \bibinfo {pages} {064017} (\bibinfo {year} {2017})},\ \Eprint
  {http://arxiv.org/abs/1707.05319} {arXiv:1707.05319 [gr-qc]} \BibitemShut
  {NoStop}%
\bibitem [{\citenamefont {Franchini}(2023)}]{Franchini:2023xhd}%
  \BibitemOpen
  \bibfield  {author} {\bibinfo {author} {\bibfnamefont {N.}~\bibnamefont
  {Franchini}},\ }\href {\doibase 10.1103/PhysRevD.108.044079} {\bibfield
  {journal} {\bibinfo  {journal} {Phys. Rev. D}\ }\textbf {\bibinfo {volume}
  {108}},\ \bibinfo {pages} {044079} (\bibinfo {year} {2023})},\ \Eprint
  {http://arxiv.org/abs/2305.19313} {arXiv:2305.19313 [gr-qc]} \BibitemShut
  {NoStop}%
\bibitem [{\citenamefont {Walker}\ and\ \citenamefont
  {Penrose}(1970)}]{Walker:1970un}%
  \BibitemOpen
  \bibfield  {author} {\bibinfo {author} {\bibfnamefont {M.}~\bibnamefont
  {Walker}}\ and\ \bibinfo {author} {\bibfnamefont {R.}~\bibnamefont
  {Penrose}},\ }\href {\doibase 10.1007/BF01649445} {\bibfield  {journal}
  {\bibinfo  {journal} {Commun. Math. Phys.}\ }\textbf {\bibinfo {volume}
  {18}},\ \bibinfo {pages} {265} (\bibinfo {year} {1970})}\BibitemShut
  {NoStop}%
%%CITATION = CMPHA,18,265;%%
\bibitem [{\citenamefont {Frolov}\ \emph {et~al.}(2017)\citenamefont {Frolov},
  \citenamefont {Krtous},\ and\ \citenamefont {Kubiznak}}]{Frolov:2017kze}%
  \BibitemOpen
  \bibfield  {author} {\bibinfo {author} {\bibfnamefont {V.~P.}\ \bibnamefont
  {Frolov}}, \bibinfo {author} {\bibfnamefont {P.}~\bibnamefont {Krtous}}, \
  and\ \bibinfo {author} {\bibfnamefont {D.}~\bibnamefont {Kubiznak}},\ }\href
  {\doibase 10.1007/s41114-017-0009-9} {\bibfield  {journal} {\bibinfo
  {journal} {Living Rev. Rel.}\ }\textbf {\bibinfo {volume} {20}},\ \bibinfo
  {pages} {6} (\bibinfo {year} {2017})},\ \Eprint
  {http://arxiv.org/abs/1705.05482} {arXiv:1705.05482 [gr-qc]} \BibitemShut
  {NoStop}%
\bibitem [{\citenamefont {Brill}\ \emph {et~al.}(1972)\citenamefont {Brill},
  \citenamefont {Chrzanowski}, \citenamefont {Martin~Pereira}, \citenamefont
  {Fackerell},\ and\ \citenamefont {Ipser}}]{Brill:1972xj}%
  \BibitemOpen
  \bibfield  {author} {\bibinfo {author} {\bibfnamefont {D.~R.}\ \bibnamefont
  {Brill}}, \bibinfo {author} {\bibfnamefont {P.~L.}\ \bibnamefont
  {Chrzanowski}}, \bibinfo {author} {\bibfnamefont {C.}~\bibnamefont
  {Martin~Pereira}}, \bibinfo {author} {\bibfnamefont {E.~D.}\ \bibnamefont
  {Fackerell}}, \ and\ \bibinfo {author} {\bibfnamefont {J.~R.}\ \bibnamefont
  {Ipser}},\ }\href {\doibase 10.1103/PhysRevD.5.1913} {\bibfield  {journal}
  {\bibinfo  {journal} {Phys. Rev. D}\ }\textbf {\bibinfo {volume} {5}},\
  \bibinfo {pages} {1913} (\bibinfo {year} {1972})}\BibitemShut {NoStop}%
\bibitem [{\citenamefont {Chandrasekhar}(1976)}]{Chandrasekhar:1976mx}%
  \BibitemOpen
  \bibfield  {author} {\bibinfo {author} {\bibfnamefont {S.}~\bibnamefont
  {Chandrasekhar}},\ }\href {\doibase DOI: 10.1098/rspa.1976.0056} {\bibfield
  {journal} {\bibinfo  {journal} {Proc. Roy. Soc. Lond.}\ }\textbf {\bibinfo
  {volume} {A349}},\ \bibinfo {pages} {1} (\bibinfo {year} {1976})}\BibitemShut
  {NoStop}%
%%CITATION = PRSLA,A349,571;%%
\bibitem [{\citenamefont {Chandrasekhar}(1985)}]{Chandrasekhar:1985kt}%
  \BibitemOpen
  \bibfield  {author} {\bibinfo {author} {\bibfnamefont {S.}~\bibnamefont
  {Chandrasekhar}},\ }in\ \href@noop {} {\emph {\bibinfo {booktitle} {{Oxford,
  UK: Clarendon (1992) 646 p., OXFORD, UK: CLARENDON (1985) 646 P.}}}}\
  (\bibinfo {year} {1985})\BibitemShut {NoStop}%
%%CITATION = INSPIRE-224457;%%
\bibitem [{\citenamefont {Torres
  Del~Castillo}(1988)}]{TorresDelCastillo:1988td}%
  \BibitemOpen
  \bibfield  {author} {\bibinfo {author} {\bibfnamefont {G.~F.}\ \bibnamefont
  {Torres Del~Castillo}},\ }\href {\doibase 10.1063/1.527993} {\bibfield
  {journal} {\bibinfo  {journal} {J. Math. Phys.}\ }\textbf {\bibinfo {volume}
  {29}},\ \bibinfo {pages} {971} (\bibinfo {year} {1988})}\BibitemShut
  {NoStop}%
%%CITATION = JMAPA,29,2078;%%
\bibitem [{\citenamefont {Lunin}(2017)}]{Lunin:2017drx}%
  \BibitemOpen
  \bibfield  {author} {\bibinfo {author} {\bibfnamefont {O.}~\bibnamefont
  {Lunin}},\ }\href {\doibase 10.1007/JHEP12(2017)138} {\bibfield  {journal}
  {\bibinfo  {journal} {JHEP}\ }\textbf {\bibinfo {volume} {12}},\ \bibinfo
  {pages} {138} (\bibinfo {year} {2017})},\ \Eprint
  {http://arxiv.org/abs/1708.06766} {arXiv:1708.06766 [hep-th]} \BibitemShut
  {NoStop}%
\bibitem [{\citenamefont {Dolan}(2019)}]{Dolan:2019hcw}%
  \BibitemOpen
  \bibfield  {author} {\bibinfo {author} {\bibfnamefont {S.~R.}\ \bibnamefont
  {Dolan}},\ }\href {\doibase 10.1103/PhysRevD.100.044044} {\bibfield
  {journal} {\bibinfo  {journal} {Phys. Rev. D}\ }\textbf {\bibinfo {volume}
  {100}},\ \bibinfo {pages} {044044} (\bibinfo {year} {2019})},\ \Eprint
  {http://arxiv.org/abs/1906.04808} {arXiv:1906.04808 [gr-qc]} \BibitemShut
  {NoStop}%
\bibitem [{Mei(8 23)}]{Mei:2023phov1}%
  \BibitemOpen
  \href@noop {} {}\bibinfo {howpublished}
  {\url{https://arxiv.org/abs/2311.18409v1}} (\bibinfo {year} {Accessed:
  2025-08-23})\BibitemShut {NoStop}%
\bibitem [{\citenamefont {Szekeres}(1965)}]{Szekeres:1965ux}%
  \BibitemOpen
  \bibfield  {author} {\bibinfo {author} {\bibfnamefont {P.}~\bibnamefont
  {Szekeres}},\ }\href {\doibase 10.1063/1.1704788} {\bibfield  {journal}
  {\bibinfo  {journal} {J. Math. Phys.}\ }\textbf {\bibinfo {volume} {6}},\
  \bibinfo {pages} {1387} (\bibinfo {year} {1965})}\BibitemShut {NoStop}%
\bibitem [{\citenamefont {Stephani}\ \emph {et~al.}(2003)\citenamefont
  {Stephani}, \citenamefont {Kramer}, \citenamefont {MacCallum}, \citenamefont
  {Hoenselaers},\ and\ \citenamefont {Herlt}}]{Stephani:2003tm}%
  \BibitemOpen
  \bibfield  {author} {\bibinfo {author} {\bibfnamefont {H.}~\bibnamefont
  {Stephani}}, \bibinfo {author} {\bibfnamefont {D.}~\bibnamefont {Kramer}},
  \bibinfo {author} {\bibfnamefont {M.~A.~H.}\ \bibnamefont {MacCallum}},
  \bibinfo {author} {\bibfnamefont {C.}~\bibnamefont {Hoenselaers}}, \ and\
  \bibinfo {author} {\bibfnamefont {E.}~\bibnamefont {Herlt}},\ }\href
  {\doibase 10.1017/CBO9780511535185} {\emph {\bibinfo {title} {{Exact
  solutions of Einstein's field equations}}}},\ Cambridge Monographs on
  Mathematical Physics\ (\bibinfo  {publisher} {Cambridge Univ. Press},\
  \bibinfo {address} {Cambridge},\ \bibinfo {year} {2003})\BibitemShut
  {NoStop}%
\end{thebibliography}%
%%%%%%%%%%%%%%%%%%%%%%%%%%%%%%%%%%%%%%%%%%%%%%%%%%%%%%%%%%%%%%%%
\end{document}